# MODELES DE CROISSANCE FRACTALE°: EPIDEMIES, EVOLUTIONS EN BIOLOGIE, ECOLOGIE, TRAFIC, ECONOMIE

Marcel Ausloos

SUPRATECS, B5, Universit e de Li'ege, B-4000 Li'ege, Belgique

R esum e°:
Divers mod'eles simples de croissance et de d ecroissance utilisent les concepts et mesures propres   la g eom etrie fractale. Des mod`eles « cin'etiques » pertinents propos'es pour des observations en physique de la mati`ere condens'ee sont rappel'es. On d'ecrit bri`evement comment les adapter, en introduisanr des degr'es de libert'e, pour reproduire des lois (en puissance) p.ex. sur la disparition (des esp`eces), les fractures (de nuages), les crashs (boursiers). Ceci permet d illustrer et de d ecrire des probl'emes scientifiques relevant de la croissance ou du contrôle d''epid'emies et autres formes de propagation d''ev`enements ou entit'es en biologie, 'ecologie, trafic, et 'economie.

Abstract
Several  simple growth and decay models use concepts and measures from fractal geometry. Kinetic models relevant to condensed matter observations arerecalled. They should be specifically adapted  with appropriate degrees of freedom in order to reproduce power law behaviors, e.g. of species extinctions, cloud fracture, financial crashes. This allows one to illustrate scientific problems relevant to the growth and control of epidemias, and other event or entity propagation in biology, ecology, traffic and economy.

1. Introduction

Divers aspects de l apport de la g eom etrie fractale (Mandelbrot, 1983, Gouyet 1996,  Sapoval 2001) peuvent ^etre mis en  evidence  sur des probl'emes  scientifiques relevant de la croissance ou du contrôle d' 'epid'emies et autres formes de propagation d' 'ev`enements ou entit'es. Dans le même esprit, des consid'erations sur le rôle des id'ees de fractalit'e, les lois d' 'echelle, ont trouv'e leur place
pour d'ecrire  des processus d' 'evolution en biologie, 'ecologie, trafic, 'economie, souvent à partir de mod`eles pertinents à des observations en physique de la mati`ere condens'ee. La toile de fond du texte suivant est donc le mot « croissance », avec son antinomie, la d'ecroissance, la disparition (des esp`eces), les fractures, les crashes (boursiers, p.ex.)

2. Mod`eles simples à semence centrale

De nombreuses consid'erations analytiques sur la diffusion et la croissance d'amas cristallins ont culmin'e r'ecemment à la suite d''etudes de l''equation de Kardar-Parisi-Zhang (1986). Cependant la plupart des recherches exp'erimentales en croissance cristalline sont analys'ees à partir de r'esultats de simulation num'erique, faisant appel à des mod`eles dits « cin'etiques » (Herrmann, 1986). Les dynamiques de croissance semblent difficilement int'egrables dans les algorithmes.

En fait le premier mod`ele simple de croissance semble ^etre celui propos'e par Eden (1958) afin de quantifier la croissance de cellules canc'ereuses. Le mod`ele consiste à poser une semence sur un r'eseau (p. ex. bi-dimensionnel), à coller à celle-ci une entit'e similaire, pour former un dim`ere, et de continuer à choisir al'eatoirement le site de croissance sur le p'erim`etre de l'amas. Même si des lacunes apparaissent dans l'amas au cours de sa croissance, celles-ci se remplissent toujours puisque le p'erim`etre est 'egalement d'efini sur les couches int'erieures. La probabilit'e de trouver un type d'amas est non-triviale ; de même pour le nombre de site de croissance ; le syst`eme est hors d''equilibre et non-ergodique; sa construction a une histoire. N'eanmoins la g'eom'etrie finale de l'amas est telle que sa dimension fractale D est simplement 'egale à 2. Il est clair qu'on peut modifier l'algorithme pour consid'erer, p.ex., des croissances à partir d'entit'es non sym'etriques, et même faire de la croissance par agglom'eration d'amas (Jullien et Botet, 1985).

Une am'elioration de ce processus de croissance fut l'œuvre de Witten et Sander (1981), inventeurs du mod`ele DLA (« diffusion limited aggregation »). Apr`es avoir plac'e une semence sur un r'eseau, une entit'e similaire lanc'ee, loin de la semence (plus tard de l'amas), parcourt le r'eseau suivant un mouvement (p.ex. Brownien) et s'attache à la semence (puis à l'amas) sur le premier site voisin disponible. On obtient des structures en fjords. La dimension fractale d'epend de type de r'eseau et de la dimensionnalit'e du syst`eme.

Afin d'introduire une dynamique de croissance, on peut consid'erer (Ausloos et coll., 1993) que l'entit'e 'el'ementaire poss`ede un degr'e de libert'e. Le plus simple est de lui attribuer un spin scalaire et de consid'erer que la croissance en un site de l'amas d'epend d'une probabilit'e de type Boltzmann, - l''energie d'ependant d'un Hamiltonien de type Ising, contenant même un champ coupl'e au degr'e de libert'e. Le nombre de voisins en interaction, et la force d'interaction, sont des param`etres de l'algorithme. L'entit'e choisit le site ayant la plus forte probabilité de croissance. On forme ainsi dans ce modèle d' « Eden Magnétique » des amas contenant des domaines. Non seulement la structure de ceux-ci peut être étudiée pour leur géométrie fractale, mais le degré de liberté étant une variable thermodynamique, des considérations habituelles, comme un diagramme de phases et des transitions ordre-désordre peuvent être envisagées. Ainsi on peut montrer (Fig.1) que le minimum de la dimension fractale de l'amas arrive pour la même valeur du couplage entre spins que l'apparition d'un moment magnétique $M$ non nul.

Pour définir la probabilité de choix su site de croissance, on peut considérer un Hamiltonien de Potts, au lieu de celui d'Ising, ainsi proposant divers degrés de liberté (Fig.2). Le type de réseau peut être modifié. L'algorithme peut être modifié pour établir une règle de relaxation, suivant une dynamique de Métropolis, etc. Les mêmes extensions peuvent s'appliquer au DLA afin de développer un modèle MDLA (Vandewalle et Ausloos, 1995a). Il est intéressant de mettre en évidence qu'on peut déduire qu'il existe un nombre fini de modes de croissance, délimités dans un diagramme de phases par les lignes d'équiprobabilité de croissance.

Les extensions les plus intéressantes de ces modèles simples consistent à introduire des pièges, réactifs ou non, mobiles ou non, de tailles et propriétés physiques pertinentes à l'étude scientifique envisagée. Il est évident que le modèle est alors susceptible de décrire une épidémie se propageant dans un milieu ayant des entités vaccinées, ou non, se réfugiant dans des cliniques ou non, mourantes ou survivants (Vandewalle et Ausloos, 1996a). Dans un autre cas, celui de la croissance de céramiques cristallines supraconductrices, dites « YBCO », on a dès lots introduit des « impuretés » réactionnelles, mobiles, suivant un processus dit UCJ (Uhlman et coll., 1964). Tenant compte de la viscosité du milieu, de son anisotropie, de la température péritectique de la réaction, de la distribution en taille des particules, … on a pu proposer une optimisation de la procédure de synthèse (Cloots et al., 1996).

   3. Modèles simples à croissance directionnellement privilégiée : trafic

La propagation du front de croissance peut se faire à partir d'une semence centrale mais aussi d'un substrat. Des variantes des modèles décrits dans la Section 2 simulent des dépositions balistiques comme pour la pluie ou la croissance de films (cf. refs. in Herrmann, 1986), ou la désintégration de ceux-ci sous impact (Ausloos et Kowalski, 1992).

Ainsi fut développé le modèle appelé « du Heysel », en souvenir du match Liverpool-Juventus du 29.05.1985. Des entités, dites « impuretés », en concentration $c$, sont distribuées aléatoirement sur des sites d'un réseau carré, borné à gauche par un substrat d'où se développe un front, un site à la fois. Les impuretés (supporters) sont mobiles et se déplacent vers la droite, p. ex. vers un des sites plus proches voisins libres, lorsqu'approch'ees par le front (hooligans). Des variantes sont possibles. On peut admettre que le front repousse les impuret es jusqu au moment o deux impuret es sont en contact, ou les d epasse. Le mod'ele est celui du trafic autoroutier en Europe ( d epassement d une impuret e, et rabattement du plus rapide, sur la bande de droite). Le cas des impuret es r eactionnelles n a pas et e etudi e. Il est quasi 'evident qu'une concentration « critique » d'impuret'es existent au dessus de laquelle le front est bloqu'e. Ce front est en fait une fractale de dimension 91/48 (Vandewalle et Ausloos, 1996b). Des lois d''echelles d'ecrivent les propri'et'es de ce front, comme la densit'e à l'avant et à l'arri`ere du front, sa largeur, le d'eplacement des particules, la taille des amas, ….

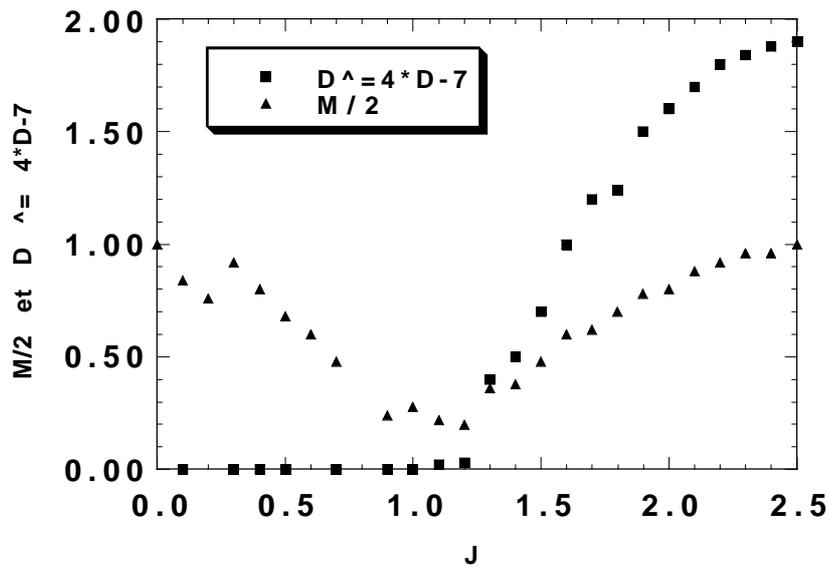

Fig. 1 L'aimantation (M) et la dimension fractale (D) d'un amas d'Eden « magn'etique » en fonction de l''energie d''echange (J) dans le Hamiltonien d'Ising repr'esentant l'interaction entre entit'es

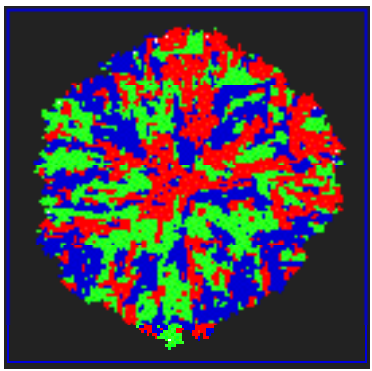

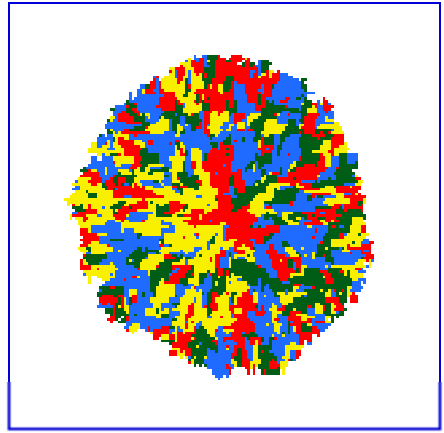

Fig. 2 Modèle de croissance d un amas d Eden ¨˚ magnétique˚¨˚; la probabilité de choix du site de croissance est déterminée par le minimum d énergie d interaction, modélisée par un modèle de Potts à 3 degrés de liberté ( droite), ou à 4 degrés de liberté ( gauche)

4. Approximations théoriques et Application écologique

Pour des considérations théoriques, de l ordre de l approximation du champ moyen, le réseau imaginé peut-être un arbre de Bethe ou Cayley sur lequel sont placées des impuretés, mobiles ou non, avec des degrés de liberté ou non pour celles-ci. L arbre, de type aussi phylogénétique, .est alors self-similaire. Trois types d arbres fractals ont été examinés (Vandewalle et Ausloos˚, 1995b, 1996c,1996d, 1997a), suivant que la longueur des branches, suit une loi d échelles, que le nombre de bourgeons (ou de branches élaguées), ou que le nombre de branchements varie. On peut également imposer une dynamique stochastique d évolution, basée sur les idées de Darwin. L arbre est développé avec un processus de branchement tenant compte d effets de compétitions et de corrélations. Des espèces peuvent se développer à chaque branchement. Celles-ci peuvent interagir entre elles, être en compétition, modifier leur capacité à survivre dans l environnement créée par les autres. Chaque entité est caractérisée par un degré de liberté, sa ¨˚fitness˚¨, évoluant au cours du temps, à la Lamarck ou à la Darwin, p.ex. Une auto organisation critique (cf. Cessac, 2004) est possible, lorsqu une dynamique de Bak-Sneppen (1993) est implémentée (Kramer et coll., 1996), à savoir la plus faible espèce est remplacée par une autre et les ¨˚fitness˚¨ des voisines modifiées.. En présence de corrélations à courte portée, le processus s organise dans un état critique caractérisé par des explosions d activités de toute taille et des intermittences. La dynamique des régimes transitoires montre que le paramètre d ordre décroit suivant une loi de puissance. La portée génétique $k$ des corrélations et compétitions entre espèces vivantes est le paramètre crucial pour déterminer la classe d universalité. P. ex. la dimension fractale varie de 2 à l infini si $k$ varie de 1 à l infini. L exposant critique de la distribution des avalanches varie de 1.5 à 1.2 lorsque $k$ augmente de 1 à 10. Si la distance d interaction $k$ (Vandewalle et coll., 1998) est de taille finie, il semble clair que certaines espèces soient écrantées de l évolution globale, et ne modifient plus leur ¨˚fitness˚¨ (Vandewalle et Ausloos˚, 1996e). La distribution de ces espèces suit une loi exponentielle. Cependant si l'interaction varie moins vite qu'en (1/r), la distribution est une loi en puissance. Le cas particulier du génocide a été aussi envisagé (Ausloos et coll., 1999).

5. Applications en économie

Le mot ¨croissance¨ a souvent une connotation économique. Le SP500, le DJIA, le NASDAQ, le GDP, sont des indicateurs de la croissance et du bien être. En outre des lois d'échelles ont été découvertes tant sur des indices que sur d'autres mesures, comme le prix du coton (Mandelbrot, 1997). Il est naturel d'avoir essayé de développer des modèles stochastiques ou microscopiques pour reproduire les évolutions de ces indicateurs, quitte à essayer même de prédire leur évolution. Ceci revient à étudier l'évolution de signaux temporels (cf. Levy-Vehel, 2004). Une caractéristique essentielle est la dimension fractale du signal. Une des méthodes rapides consiste à étudier l'exposant de Hurst, H, relié simplement à $D$, $D=2-H$, et l'exposant de la puissance spectrale $\beta$. La technique de ¨detrended fluctuation analysis¨ (DFA) est alors très utile (Ausloos, 2000, Ausloos et coll., 2000). Comme ces exposants peuvent être définis dans des fenêtres de taille finie, on parle d'exposants locaux. Leur distribution peut démontrer une structure multifractale (cf. Arneodo, 2004), - d'autant plus que les marchés financiers sont souvent dits ¨turbulents¨.

La connaissance d'un $D$ local permet en fait d'interpréter les fluctuations du signal comme étant corrélées ou anticorrélées. Ceci a conduit à élaborer des stratégies d'investissement basées sur des modèles simples, puisqu'on peut prévoir avec une certaine probabilité le type (signe et amplitude) de fluctuations. Des gains théoriques importants sont ainsi possibles (Vandewalle et Ausloos, 1997b). De plus, on peut percevoir *a posteriori* l'influence de spéculateurs ou observer le délit d'initié. En effet, il apparaît que les acteurs contrôlant les politiques économiques ¨souhaitent¨ que les fluctuations soient gaussiennes, en d'autres termes que le signal soit de caractéristique Brownienne, soit $D=1.5$. Les modèles microscopiques actuels (Ausloos et al., 2003, 2004) essayent d'intégrer cette structure en utilisant des entités contenant l'un ou l'autre degré de liberté et réagissant à l'un ou l'autre ¨champ (extérieur) économique¨.

Incidemment on a constaté que le même processus se retrouve dans des signaux météorologiques, comme la hauteur de la base des nuages. Celle-ci fluctue de manière anti-corrélée, jusqu'au moment où les fluctuations tendent vers celle d'un mouvement Brownien (Ivanova et Ausloos, 1999, Ivanova et coll., 2000). Le même type de modèle de (dé)croissance s'applique pour les nuages et les cours de la bourse. Il s'agit bien en effet de turbulence !

D'autre part, certains crashes boursiers sont caractérisés par des oscillations log-périodiques dans la bulle précédant le crash (Vandewalle et coll., 1999). Ces oscillations correspondent à la partie imaginaire de la dimension fractale, caractéristique d'événements de croissance sur des systèmes à structure d'invariance discrète (Sornette, 1998). Le modèle du tas de sable sur base fractale (Ausloos et coll., 2002) est un paradigme de ce phénomène. Il est facilement démontré analytiquement et par simulation que cela se produit sur des arbres ou réseaux à géométrie fractale, p.ex. l'arbre ou la carpette de Sierpinski.

L analogie avec des phénomènes de dimension plus élevées, comme les surfaces fracturées (Ausloos et Berman, 1985) et les tremblements de terre (cf. Schertzer, 2004), est d un intérêt évident pour des travaux futurs.

6. Conclusions

De simples modèles cinétiques de croissance, et de décroissance, bien adaptés pour introduire l un ou l autre degré de liberté, couplé ou non à un champ extérieur, sont disponibles pour étudier, décrire, prédire le comportement de signaux spatio-temporels, caractéristiques d évolutions (fractales) en biologie, écologie, trafic, économie. Ces modèles trouvent leurs sources dans des problèmes de matière condensée°: la croissance de cristaux, ou les fractures de matériaux.